\documentclass[epj,twocolumn]{svjour}
\usepackage{amsmath,amsfonts,amssymb,graphics,color,bm}
\usepackage{hyperref}
\usepackage{ulem}

\newcommand{\be}{\begin{equation}}
\newcommand{\ee}{\end{equation}}
\newcommand{\bea}{\begin{eqnarray}}
\newcommand{\eea}{\end{eqnarray}}

\newcommand{\hf}{\hat{ \varphi}}
\newcommand{\hvf}{\hat{\bm \varphi}}

 \def\CL{{\cal{L}}}
 \def\be{\begin{equation}}
 \def\ee{\end{equation}}

 \def\m{\mu}

 \def\sl{\ds}
 \def\ds#1{#1\kern-1ex\hbox{/}}
 \def\sla{\raise.15ex\hbox{$/$}\kern-.57em}

 \def\({\left(}
 \def\){\right)}
 \def\[{\left[}
 \def\]{\right]}
 
\newcommand{\Eq}[1]{Eq.~(\ref{#1})}

\begin{document}
\title{Scrutinizing the pion condensed phase}
\author{Stefano Carignano\inst{1} \and  Luca Lepori\inst{2,3} \and Andrea Mammarella\inst{1} \and Massimo Mannarelli\inst{1} \and Giulia Pagliaroli\inst{1,4}}
\institute{INFN, Laboratori Nazionali del Gran Sasso, Via G. Acitelli, 22, I-67100 Assergi (AQ), Italy \and Dipartimento di Fisica e Astronomia, Universit\'a di Padova, Via Marzolo 8, I-35131 Padova, Italy \and Dipartimento di Scienze Fisiche e Chimiche, Universit\'a dell'Aquila, via Vetoio,
I-67010 Coppito-L'Aquila, Italy \and Gran Sasso Science Institute, Viale Francesco Crispi, 7,  L'Aquila, Italy}
\date{Received: date / Revised version: date}
\abstract{
When the isospin chemical potential exceeds the pion mass, charged pions condense in the zero-momentum state forming a superfluid. Chiral perturbation theory provides a very powerful tool for studying this phase. However, the formalism that is usually employed in this context does not clarify various aspects of the condensation mechanism and makes the identification of the soft modes problematic. We re-examine the pion condensed phase using different approaches within the chiral perturbation theory framework. As a first step, we perform a low-density expansion of the chiral Lagrangian valid  close to the onset of the Bose-Einstein condensation. We obtain an effective theory that can be mapped to a Gross-Pitaevskii Lagrangian in which, remarkably, all the coefficients depend on the isospin chemical potential. The low-density expansion becomes unreliable deep in the pion condensed phase. For this reason, we develop an alternative field expansion deriving a low-energy Lagrangian analog to that of quantum magnets. By integrating out the ``radial" fluctuations we obtain a soft Lagrangian in terms of the Nambu-Goldstone bosons arising from the breaking of the pion number symmetry. Finally, we test the robustness of the second-order transition between the normal and the pion condensed phase when next-to-leading-order chiral corrections are included. We determine the range of parameters for turning the second-order phase transition into a first-order one, finding that the  currently accepted values of these corrections are unlikely to change the order of the phase transition.
\PACS{
      {11.30.Rd}{Chiral symmetries}   \and
      {12.39.Fe}{Chiral Lagrangians} \and
      {67.25.D}{Superfluid phase}
     } 
} 
\maketitle
\section{Introduction}

Systems with a nonvanishing isospin chemical potential, $\mu_I$, are very good playgrounds for gaining insight on  quantum chromodynamics (QCD) in the nonperturbative regime. { Indeed, several complementary approaches can be employed to study them, possibly leading to a more solid understanding of their properties.}  The first results on the properties  of matter at vanishing temperature as a function  of  $\mu_I$ have been obtained by chiral perturbation theory ($\chi$PT), see~\cite{Son:2000xc,Kogut:2001id}. At not too large $\mu_I$, lattice QCD simulations of pions (and kaons) are feasible and have been developed in~\cite{Alford:1998sd,Kogut:2002zg,Detmold:2012wc,Detmold:2008yn,Cea:2012ev,Endrodi:2014lja}. The Nambu-Jona Lasinio (NJL) models can be used in a wide range of the isospin chemical potential~\cite{Toublan:2003tt,Barducci:2004tt,Barducci:2004nc,Ebert:2005wr,Ebert:2005cs,He:2005nk,He:2010nb}  and random matrix models have been developed as well~\cite{Klein:2004hv,Kanazawa:2014lga}. Perturbative methods have been used in~\cite{Graf:2015pyl} for addressing a region outside the realm of  $\chi$PT, but possibly reachable by future lattice QCD simulations. Finite temperature effects were also considered in several  works~\cite{He:2010nb,Loewe:2002tw,Loewe:2004mu,Xia:2014bla,Loewe:2016wsk}.
 Non-homogeneous phases at finite isospin densities have been studied in~\cite{Ebert:2011rg,Carignano:2015kda,Nowakowski:2015ksa,Adhikari:2016vuu}.  Clearly, all of these models have a drawback. For example,
present lattice simulations at $\mu_I \gtrsim 2 m_\pi$ have large  errors and  NJL model results depend on the parameter sets employed.

The isospin chemical potential has several effects on hadronic matter. Not only it induces a  Zeeman-like mass splitting  within isomultiplets, but it can also  rotate  the quark-antiquark condensate. The latter phenomenon is called pion condensation, because it is characterized by the breaking of the $U(1)$ global symmetry corresponding to the conservation of the pion number and the occurrence of a Bose-Einstein condensate (BEC) of pions.  Since condensed pions are charged, the system is  a superfluid of charged particles, that is, an electromagnetic superconductor~\cite{Mammarella:2015pxa}. 
Using the leading order (LO) $\chi$PT Lagrangian it was found that at  $\mu_I=m_\pi$  a second order phase transition between 
the normal  and  the pion-condensed ($\pi c$) phase occurs~\cite{Son:2000xc}.

In this work we will use a realization of    $\chi$PT that includes only pions~\cite{Gasser:1983yg,Leutwyler:1993iq,Ecker:1994gg,Scherer:2002tk}, meaning that we will restrict ourselves to $|\mu_I| < m_\rho \sim 770$ MeV. We also assume that the strange quark chemical potential is so small that it does not allow the appearance of kaons. The present analysis can be easily extended to the kaon condensed phase, see for example the discussion in~\cite{Carignano:2016rvs}. 
Since we only deal with mesons, the baryonic chemical potential  $\mu_B$ does not appear in the Lagrangian.  We shall implicitly assume that  $\mu_B$ is  below the nucleon mass, see for example the discussion in~\cite{Kogut:2001id} and in~\cite{Mammarella:2015pxa}.

From the $\chi$PT perspective,  the possibility of comparing its results with those obtained using different methods 
would allow for a  tuning of its parameters and lead to a better understanding of the ground state and the  low-energy properties of the system. As a remarkable example, in the lattice QCD simulations of~\cite{Detmold:2012wc} it was found that the ratio between the energy density and the Stefan-Boltzmann energy density  has a  peak  at  $\mu_I^\text{peak} \simeq 1.27 m_\pi$. In~\cite{Carignano:2016rvs} it has been shown that this peak structure can be accurately described using the  LO $\chi$PT Lagrangian, with  an analytic result for the peak position $\mu_I^\text{peak} = ({\sqrt{13} -2}\, )^{1/2} m_\pi \simeq 1.27\, m_\pi$. This peak seems to be related with the saturation of the pion condensate. Moreover, we have indications from  the  LO $\chi$PT Lagrangian that the system should make a smooth crossover from the  BEC phase to a Bardeen-Cooper-Schrieffer (BCS) phase at $\bar\mu_I \simeq \sqrt{3} m_\pi$, in agreement with the  NJL findings of~\cite{He:2010nb}, giving  $\bar\mu_I \sim 1.6-2\, m_\pi$, depending on the parameter set considered.

Although the $\chi$PT expansion  is extremely powerful for the determination of the ground state properties and of the low-lying excited states,  the physical interpretation of the $\pi c$ phase is, in our opinion, not so clear. One of the reasons is that for $\mu_I =0$ the charged pions are charge conjugate fields. Turning on  the isospin chemical potential explicitly breaks this symmetry, therefore the identification of the physical states is cumbersome. The situation becomes even more complicated in the $\pi c$ phase, in which the Lagrangian takes a non trivial expression and the low-lying states are given by complicated combinations of the pion fields, see for example~\cite{Mammarella:2015pxa}.

In order to shed some light on the $\pi$c phase, we will  provide different expansions of  the  $\chi$PT Lagrangian, aiming at a more accessible  physical interpretation of known results obtained at $T=0$. In particular, we will  rewrite the Lagrangian in forms similar to those obtained in condensed matter systems or  in the study of superfluid phenomena. In the context of weakly interacting bosonic systems, various different approaches can be used, see~\cite{Andersen:2003qj} for a review.   Close to the BEC onset  we derive a low-density expansion having exactly the form of a Gross-Pitaevskii (GP) equation. This approach is similar to the one  developed in  standard  dilute   bosonic systems~\cite{Andersen:2003qj}  considering a  $\sqrt{n a^3}$ expansion, where $n$ is the number density and $a$ is the $s$-wave scattering length.   We name it the low-density expansion because, as we will see, the tree-level $2\to 2$ scattering amplitude tends to a constant nonvanishing value at the BEC phase transition point. Therefore, the actual control parameter in the $\pi$c phase is the density of pions in the condensate. 
Remarkably,  both the effective chemical potential, the effective mass and the $2\to 2$ scattering amplitude of the GP Lagrangian obtained by the low-density expansion depend  on $\mu_I$. This identification clarifies one of the reasons why the description of the $\pi c$ phase is  complicated: by changing $\mu_I$, all of these quantities simultaneously change. For this part of the discussion we will restrict ourselves to leading-order results, but next-to-leading order terms in the low-density expansion can be straightforwardly determined.

For larger values of $\mu_I$, the low-density expansion breaks down. The basic reason is that the system is no more dilute, with a large number of pions occupying the ground state. 
In order to gain insight on the properties of the system we provide an alternative  description of the $\pi c$ phase, similar to the low-energy description of quantum magnets.  We identify two different excitations: the radial, or Higgs, mode  corresponding to amplitude fluctuations of the pion condensate, and the  phase oscillation of the condensate corresponding to the massless Nambu-Goldstone boson (NGB). This mode is associated with the breaking of the global  $U(1)$ symmetry related to the  pion number conservation and is also known in condensed matter physics as the Anderson-Bogoliubov mode; it can also be interpreted as the quasiparticle associated with the propagation of pressure perturbations, thus we sometimes call it the phonon.  We find a soft Lagrangian similar to the Heisenberg model for quantum magnets in which isospin plays the role of spin in condensed matter. Therefore, the pion condensed phase can be thought as an ordered magnetic phase with isospin aligned along the $\mu_I$ direction.
{{By integrating}  out the radial fluctuations we derive all the interaction terms and the surface terms of the soft Lagrangian. In particular, since the isospin chemical potential explicitly breaks the Lorentz boost invariance but not the rotation invariance, the kinetic term and the interaction terms can be written using an analogue model of gravity.

Finally, we scrutinize the effect of the next-to-leading-order (NLO) chiral corrections in the static Lagrangian. We examine, for the first time, the effect of these corrections on the phase transition between the normal  and the pion condensed phases. {After} including these corrections,  the phase transition happens at values of $\mu_I$ of the order of the NLO pion mass.   We also determine the range of the NLO { low-energy constants} necessary for changing the order of the  transition between the normal  and the $\pi$c phases.
 
The present paper is organized as follows. In Sec.~\ref{sec:summary} we summarize some of the most interesting properties of   the $\pi$c phase  at vanishing temperature obtained in the literature.  In Sec.~\ref{sec:low-energy} we present a low-energy effective theory valid close to the normal phase-BEC phase transition.  In Sec.~\ref{sec:alternative} we provide an alternative description of the $\pi c$ phase valid within the $\chi$PT validity range.  In Sec.~\ref{sec:NLO} we consider the effect of the NLO  corrections in the static Lagrangian. We draw our conclusions in Sec.~\ref{sec:conclusion}. In the Appendix A we speculate on the existence of self-bound pion stars. In the Appendix B we clarify  some aspects of the alternative procedure developed in Sec.~\ref{sec:alternative}.

\section{Brief summary of the standard $\chi$PT  results on  the pion-condensed phase}\label{sec:summary}
The $ {\cal O}(p^2)$  $\chi$PT  Lorentz-invariant Lagrangian describing the interaction of pions with an external vector field, $v_\mu$, can be written as follows~\cite{Kogut:2001id,Gasser:1983yg,Scherer:2002tk}
\begin{align}
\label{eq:Lagrangian1}
{\cal L}_2 = \frac{f_\pi^2}{4} \text{Tr} (D_\nu \Sigma D^\nu \Sigma^\dagger) + \frac{f_\pi^2 m_\pi^2}{4} \text{Tr} (\Sigma + \Sigma^\dagger )\,,
\end{align}
where \be
D_\mu \Sigma = \partial_\mu\Sigma - \frac{i}{2} [v_\mu,\Sigma] \,
\ee
is the covariant derivative and
 $\Sigma$ is an $SU(2)$ matrix collecting the pion fields. Any explicit expression for $\Sigma$ will correspond to a given parameterization of the pion fields, although
  results for physical observables are independent of the parameterization of $\Sigma$ used~\cite{Chisholm:1961tha,Kamefuchi:1961sb,Weinberg:1978kz}. 
  We will exploit this freedom to  develop different expressions of the chiral Lagrangian. 
  In the standard description, see for example~\cite{Kogut:2001id},  the  pion degrees of freedom are introduced by considering  
\be\label{eq:sigma}
\Sigma= u \bar \Sigma u \qquad \text{with} \qquad u=e^{i  \bm \sigma \cdot \bm \varphi/2} \,,
\ee
where $\varphi_i$ ($i=1,2,3$)  are real scalar fields, $\sigma_i$ are the  Pauli matrices and 
\be\label{eq:barsigma}
\bar\Sigma = e^{i \bm\alpha\cdot\bm\sigma } = \cos \alpha + i \bm n\cdot\bm\sigma\sin \alpha\,,
\ee
is the most general  $SU(2)$ vacuum, with $\alpha$ and $\bm n$ variational parameters  to be  determined by maximizing the static Lagrangian. The   low-energy constants (LECs)  $f_\pi$ and  $m_\pi$ in Eq.~\eqref{eq:Lagrangian1} correspond to the pion decay constant and  to  the pion mass, respectively. These LECs  can be related to microscopic quantities of the underlying quark model, see for example~\cite{Gasser:1983yg,Leutwyler:1993iq,Ecker:1994gg,Scherer:2002tk}.  We assume that all three pions have exactly the same mass at vanishing chemical potential.   The isospin chemical potential can be introduced as the time component of the external vector field, that is
\be\label{eq:vmu}
 v_\mu= \mu_I \delta_{\m 0} \sigma_3\,,
\ee
and it is considered as a tunable parameter in the grand-canonical approach. Clearly, it breaks Lorentz-boost invariance and  isospin symmetry, because it indicates a privileged  reference frame as well as a direction in isospin space. 

In the normal phase $\alpha=0$, meaning that $\bar \Sigma = I$ and the only effect of introducing $\mu_I$ is a Zeeman-like energy splitting proportional to the isospin charge.
Thus, the charged pion fields corresponding to
\be\label{eq:pions}
\pi_{\pm}=\frac{\varphi_1\mp i \varphi_2}{\sqrt{2}}\,,
\ee
have effective masses
\be
m_{\pi_\pm} = m_\pi \mp \mu_I\,, 
\ee
and since the neutral pion field $\pi^0 = \varphi_3$ has vanishing isospin, it follows that   $m_{\pi^0}=m_\pi$. In the following we will assume for definiteness $\mu_I \ge0$
and  we will use the adimensional quantity
 \be
 \label{eq:gamma}
  \gamma=\frac{\mu_I}{m_\pi}\,,
   \ee
as control parameter to characterize the strength of the isospin chemical potential. 

In principle, a system of pions decays into leptons.  Therefore, in order to perform a study like ours, one typically neglects electroweak interactions (usually by considering time scales much shorter than their characteristic one). 
However, let us emphasize that for $\mu_I > m_\pi -m_e $, where $m_e$ is the electron mass,  the effective  $m_{\pi_+}$ is so small that it cannot decay into leptons~\cite{Mammarella:2015pxa}. The corresponding pion number, $N_{\pi_+}$, is thus conserved even when including electroweak interactions. However, at  $\gamma=1$  the $\pi_+$ becomes massless and  the $U(1)$ symmetry corresponding to  $N_{\pi_+}$ conservation is spontaneously broken: the system becomes a superfluid.   Since the condensed bosons are electrically charged, the resulting phase is actually a superconductor~\cite{Mammarella:2015pxa}.  This symmetry breaking mechanism can be described by maximizing the ${\cal O}(p^2)$  ground state Lagrangian. 
For $\gamma > 1$, the  energetically favored phase is characterized by
$\cos \alpha=1/\gamma^2$, meaning that  the large isospin chemical potential has changed the property of the vacuum. 

The identification of the ground state having $\alpha\neq 0$  with a superfluid is based on several facts that we briefly review. It is possible to show that the chiral condensate is rotated to a  pion condensate, more specifically~\cite{Kogut:2001id} 
\begin{align}
\langle \bar u u \rangle = \langle \bar d d \rangle   \propto  \cos\alpha\,, \label{eq:chiral_cond}\\
\langle \bar d \gamma_5 u + \text{h.c.} \rangle \propto \sin\alpha\,, \label{eq:pion_cond}
\end{align}
where $u$ and $d$ are up and down quarks respectively (color and spinorial indices have been suppressed).
A nonvanishing pion condensate implies that the vacuum does not annihilate the isospin charge. The condensed  pions contribute to the total pressure and density of the system. Since we are considering vanishing temperatures, these contributions are due to the occupation of the zero energy state by a macroscopic number of particles.  The  normalized pressure (obtained subtracting the vacuum pressure) and the  number density are respectively  given by~\cite{Son:2000xc,Kogut:2001id} 
\be
P = \frac{f_\pi^2 m_\pi^2}{2} \gamma^2 \left(1-\frac{1}{\gamma^2}\right)^2,\qquad  n_{I}= f_\pi^2 m_\pi \gamma \(1-\frac{1}{\gamma^4}\)\,, \label{eq:pressure-density}
\ee
leading to the ${\cal O}(p^2)$ equation of state~\cite{Carignano:2016rvs}
\be
\epsilon(P) = -P + 2\sqrt{P(2f_\pi^2 m_\pi^2+P)}  \label{eq:eq-state}\,.
\ee
The isospin number density, which is equivalent to the electric charge one, exactly corresponds to the number density of particles in the ground state. 

Regarding the excitations, it can be shown that there exists a flat direction of the potential, which is  a typical feature of the BEC phase because it is associated with the existence of NGBs. Indeed,  by a variational procedure  it is possible to show that the unit vector $\bm n$ in Eq.~\eqref{eq:barsigma} has to be orthogonal to the direction taken by the vector field $v_\mu$ in isospin space~\cite{Mammarella:2015pxa}. This residual $O(2)$ isospin symmetry  corresponds to the flat direction of the potential.  The low-energy excitations  are given by two orthogonal combinations of the pion fields; the corresponding  dispersion laws can be found in~\cite{Mammarella:2015pxa}.  The dispersion law of the NGB mode  is given by
\be
E = \frac{m_\pi}{\sqrt{2} \gamma}\sqrt{3 + \gamma^4 + 2 p^2 \frac{ \gamma^2}{m_\pi^2} -\sqrt{(3 + \gamma^4)^2+16 p^2 \frac{\gamma^2}{m_\pi^2}}}\,, 
\ee
which can be expanded in two different regimes:
\begin{align}
E &= \frac{p^2}{2 m_\pi}  &\text{for  } \gamma=1 \label{eq:quadratic}\,,\\
E &= c_s p  + {\cal O}(p)^3   &\text{for  } \gamma>1\,,\label{eq:linear}
\end{align}
where \be\label{eq:cs}
c_s = \sqrt{ \frac{\gamma^4-1}{\gamma^4+3}} \,,
\ee
 is the sound speed. The above results indicate  that this is indeed a gapless mode. However, at the phase transition this  mode seems to interpolate between two different excitations. {Indeed, } for $\gamma=1$ it has  a quadratic dispersion law  and thus describes the $\pi_+$  that becomes massless at the phase transition point. For $\gamma>1$, it describes a mode propagating with the speed of sound (which can as well be obtained from the equation of state, Eq.~\eqref{eq:eq-state}) and should therefore correspond to the phonon. Technically, this interpolation is possible because the sound speed vanishes at the phase transition point.  

Although the above discussion is formally correct, it does not actually illustrate the symmetry breaking mechanism in detail. As we will see, the procedure for obtaining the ground state includes the nonperturbative interactions of the ${\cal O}(p^2)$ Lagrangian and this is one of the reasons why  the result looks cumbersome. Moreover, the soft Lagrangian for the NGB is difficult to identify, because the massless mode is a combination of the two charged ones. Finally,  in the standard description of the  broken phase there is still a massive mode, which should be integrated out in order to obtain the soft Lagrangian.
In the following section we will discuss the symmetry breaking mechanism using a low-energy expansion, valid in the normal phase and close to the BEC phase transition, which has a more intuitive interpretation. In Sec.~\ref{sec:alternative} we present a method for obtaining the soft Lagrangian for any $\gamma $ within the $\chi$PT  range.

\section{Low energy description of the normal phase-BEC phase transition}\label{sec:low-energy}
The $U(1)$ symmetry breaking corresponding to the violation of the pion number conservation can be described by considering the standard expression of the chiral fields
\be
\Sigma = e^{i  \bm \sigma \cdot \bm \varphi}\,,
\ee
but assuming a nonzero vacuum expectation value (vev) of one of the charged fields. Since the $\pi_0$ will not play any role, we restrict our analysis to the charged fields, meaning that we will only consider the $\varphi_1$ and $\varphi_2$ components. By expanding  the ${\cal O}(p^2)$ Lagrangian { in Eq.~\eqref{eq:Lagrangian1}} including terms up to $\varphi^4$, we find  
\begin{align}
{\cal L} &=f_\pi^2 m_\pi^2   +  i f_\pi^2 \mu_I(\pi_-\partial_0 \pi_+ -\pi_+\partial_0 \pi_-)+ f_\pi^2  \left( \partial_\nu \pi_+ \partial^\nu \pi_-\right)\nonumber \\ & -f_\pi^2(m_\pi^2 -\mu_I^2)\pi_+\pi_-+\frac{f_\pi^2}{6} (m_\pi^2-4\mu_I^2) (\pi_+\pi_-)^2 + \dots
\label{eq:lagrangian-phi4}
\end{align}
where  the pion fields have the same expression reported in Eq.~\eqref{eq:pions} and we have neglected derivative terms   ${\cal O} (\varphi^3)$ and higher. The presence of terms coupling $\pi_+$ and $\pi_-$ fields makes the study of the Lagrangian  complicated. However,  close to the phase transition it is possible to
consider excitations at arbitrarily small energies  because a massless mode exists, see also Eq.~\eqref{eq:epsilon} below. Therefore, in this region,  we can restrict the analysis to  the terms with the lowest derivative power, leading to a further simplified  soft-pion Lagrangian 
\begin{align}
{\cal L} &=f_\pi^2 m_\pi^2  +  i f_\pi^2 \mu_I(\pi_-\partial_0 \pi_+ -\pi_+\partial_0 \pi_-)- f_\pi^2  \left( {\bm \nabla} \pi_+ \cdot {\bm \nabla} \pi_-\right)\nonumber \\ & -f_\pi^2(m_\pi^2 -\mu_I^2)\pi_+\pi_-+\frac{f_\pi^2}{6} (m_\pi^2-4\mu_I^2) (\pi_+\pi_-)^2\,.
\label{eq:low-energy-lagrangian}
\end{align}
From this, one obtains the following
 equations of motion for the $\pi_\pm$ fields: 
\begin{align}
\pm 2 i f_\pi^2 \mu_I \partial_0 \pi_{\mp} &=  - f_\pi^2 \nabla^2 \pi_\mp + f_\pi^2 (m_\pi^2-\mu_I^2) \pi_\mp \nonumber \\ &- \frac{1}{3} f_\pi^2 (m_\pi^2-4 \mu_I^2) (\pi_+ \pi_-) \pi_\mp \,,
\end{align}
showing that $\pi_+$ corresponds to the low-energy particle state and $\pi_-$ to the low-energy antiparticle one.  As already noted in~\cite{Schafer:2001bq},  there is only one independent  degree of freedom, because in the low-energy limit $\pi_+$ and $\pi_-$ are conjugate fields. For any $\mu_I > 0$ we can  define
\be
\psi = \sqrt{2 f_\pi^2 \mu_I}  \, \pi_+ \,,
\ee
 thus Eq.~\eqref{eq:low-energy-lagrangian} takes the form of  a Gross-Pitaevskii (GP) Lagrangian
\be\label{eq:GP-lagrangian}
{\cal L}_\text{GP} = f_\pi^2 m_\pi^2 + i \psi^* \partial_0 \psi + \mu_\text{eff} \,  \psi^* \psi -\frac{g}2 \vert \psi^* \psi \vert^2 +\psi^* \frac{\nabla^2}{2 M} \psi\,,
\ee
with
\be \label{eq:GPcoeffs}
 \mu_\text{eff}=\frac{\mu_I^2-m_\pi^2}{2 \mu_I}\,,\qquad g =\frac{4 \mu_I^2-m_\pi^2}{12 f_\pi^2  \mu_I^2}\,, \qquad M=\mu_I\,,
 \ee
the relevant GP coefficients. It is quite striking to see the very nontrivial effect of the isospin chemical potential. Not only it changes the effective chemical potential, but it also changes the boson-boson coupling  and the coefficient of the Laplacian operator. { The effective coupling constant is related to the $\pi^+\pi^+$ scattering length, $a$, by the standard GP relation $a \propto g\, m_\pi$. However,  the  scattering  amplitude of different pions cannot be simply determined in this way because isospin is explicitly broken in the Lagrangian, see Eqs.~\eqref{eq:Lagrangian1} and~\eqref{eq:vmu}. In particular, no simple relation exists between the scattering amplitudes in the various isospin channels.}

Since the  mode is bosonic,  the effective chemical potential in the unbroken phase must be nonpositive, and indeed this happens for $\mu_I \le m_\pi$.
 Since we are working at $T=0$,  the region $\mu_I<m_\pi$ corresponds to the vacuum with no pions, whereas   for $\mu_I=m_\pi$ pions appear and the effective chemical potential vanishes. Whether this condition corresponds to the onset of the BEC regime  depends on the sign of the effective interaction. For attractive interactions the system should collapse, while for repulsive ones it becomes superfluid. 
 {From our analysis, \Eq{eq:GPcoeffs} indicates that } the interaction coupling becomes positive for  $\mu_I > m_\pi/2$, thus  the interaction between the $\psi$ fields is repulsive and the system is expected to turn into a BEC when the effective chemical potential vanishes. It is quite remarkable that well before reaching the transition point, the interaction turns from attractive to  repulsive. It might be of interest to study within this GP model what happens at a small nonvanishing temperature for $\mu_I < m_\pi/2$, corresponding to  a system in which few thermal excitations interact  with a negative scattering length. In this case the GP Lagrangian in Eq.\eqref{eq:GP-lagrangian} is still valid because the neglected derivative interactions   are thermally suppressed. This approximation is expected to break down only close to  $\mu_I = m_\pi/2$, when $g$ vanishes.} We postpone this study to future work.

From Eq.~\eqref{eq:GP-lagrangian}, we have the potential
\be
V(n)=  \mu_\text{eff}\, n - \frac{g}2 n^2\,,
\ee 
where $n=\psi^* \psi$.  The ground state number density is obtained from $\partial V/\partial n =0$, leading to
\be\label{eq:n_0}
n=6 f_\pi^2 m_\pi \gamma \frac{\gamma^2-1}{4\gamma^2-1} \,,
\ee
where $\gamma$ has been defined in Eq.~\eqref{eq:gamma}. Close to the BEC phase transition we can write 
\be\label{eq:expansion} 
\gamma=1+\varepsilon\,\ee 
with $\varepsilon \ll1$, and we obtain the approximate expression
\be\label{eq:n_0_app}
n= 4 \varepsilon f_\pi^2 m_\pi\,.
\ee 
The adimensional parameter $\varepsilon$ in Eq. \eqref{eq:expansion}  corresponds in cold atoms experiments to the quantity $x = n \, a^3$, with $a$ being the s-wave scattering length of the bosonic  atoms or molecules involved in the condensate \cite{dalfovo1999}. In particular, the condition $x \ll 1$ assures the validity of the GP description. Since the neglected terms in Eq.~\eqref{eq:lagrangian-phi4} are of order $(\psi^* \psi)^3\propto \epsilon^3$, it follows that the obtained approximation is only valid at the leading order in $\epsilon$.
Indeed, the ground state  number density in the broken phase is given by Eq.~\eqref{eq:pressure-density},
which agrees with Eq.~\eqref{eq:n_0_app} at the leading order in $\varepsilon$. This expansion is therefore a low-density expansion, and is expected to breakdown for large $n$, or more precisely, for $\varepsilon $ of order unity. 
Physically, the low-density expansion makes sense because close to the $\pi$c phase transition the number density of particles  is arbitrarily small, thus interactions involving higher order terms are suppressed.

Close to the phase transition point, the normalized pressure  is given by
\be
P= \frac{g}2 n^2=2 \varepsilon^2 f_\pi^2 m_\pi^2  \,,
\ee
which agrees with the result reported in Eq.~\eqref{eq:pressure-density}  at the leading order in $\varepsilon$. 
We thus conclude  that  close to the normal phase-BEC phase transition, the system can be approximated by the low-density expansion of the chiral Lagrangian that can be mapped to the GP Lagrangian of Eq.~\eqref{eq:GP-lagrangian} having
\be
\mu_\text{eff} = \varepsilon\,, \qquad g= \frac{1}{4 f_\pi^2}\(1+\frac{3}{2} \varepsilon\)\,, \qquad  M= m_\pi(1+\varepsilon)\,, \ee
and the resulting  pressure and ground state number density are respectively
\be
P = 2 \varepsilon^2 f_\pi^2 m_\pi^2\,, \qquad n= 4 \varepsilon f_\pi^2 m_\pi\,.
\ee

Regarding the excitations, the equation of motion takes the form of a GP equation 
\be
i \partial_0 \psi = -\frac{\nabla^2}{2 \mu_I} \psi - \mu_\text{eff} \, \psi + g \vert \psi \vert^2\psi\,,
\ee
thus, neglecting interactions,  the  dispersion law of the $\psi$ mode is given by
\be\label{eq:epsilon}
\epsilon(p)= \frac{p^2}{2 \mu_I} -\mu_\text{eff}\,,
\ee
which already includes the effect of the effective chemical potential. Thus, the  mode is gapless with a quadratic dispersion law at $\gamma=1$, matching the result obtained in the previous section, see Eq.~\eqref{eq:quadratic}, if one expands for small $\epsilon$. Within the GP framework it is also possible to obtain and to better understand the linear dispersion mode given in Eq.~\eqref{eq:linear}.
Since in the broken phase the interaction is repulsive, the low-energy excitation is a  Bogolyubov mode, or NGB, with dispersion law
\be
\epsilon_\text{NGB}(p)=p \sqrt{\frac{n g}{\mu_I}}= p \sqrt{\varepsilon} \,,
\ee
which agrees with the result reported in  Eq.~\eqref{eq:cs} considering the expansion in Eq.~\eqref{eq:expansion}.

The two different limits of the dispersion law  reported in Eqs.~\eqref{eq:quadratic} and \eqref{eq:linear}, do actually correspond to different modes: the first one is the quasiparticle mode becoming massless at the phase transition point, while the second one is the long-wavelength fluctuations corresponding to longitudinal compression mode, which exists because the boson-boson interaction is repulsive. 

The second-order phase transition can therefore be described by the standard GP Lagrangian for sufficiently small values of $\varepsilon$. It is maybe of interest the fact that at the leading order the effective coupling depends only on $1/f_\pi^2$. Moreover, the strength of the interaction increases with  $\mu_I$. This is the expected behavior for a   system  evolving towards a BCS phase. Unfortunately, the BEC-BCS crossover is expected to happen at $\gamma\simeq\sqrt{3}$, see~\cite{Carignano:2016rvs}, and it is therefore outside the range of the low-density expansion.

\section{Alternative description of the pion condensed  phase}\label{sec:alternative}
In the previous section we have clarified that  the pion condensation mechanism  can be better understood by a mapping of the chiral Lagrangian into a GP Lagrangian, which however should include an infinite number of terms for $\mu_I$ much larger than $m_\pi$. Therefore, we should find an alternative  way for treating the system in that regime. In this section we  present an  approach to the pion condensed phase aiming at a soft Lagrangian density valid for momenta below $\mu_I$. We expand {our Lagrangian} close to the potential minimum  to identify the flat direction of the potential. The important aspect is that in the broken phase, where $\gamma>1$, there still exists a massive degree of freedom corresponding to the {\it radial} fluctuation of the condensate, which is the so-called Higgs mode. In order to determine the correct expression of the soft Lagrangian we have to integrate out this mode. Clearly, one has first to identify it. 

Let us use the following field definition
\be\label{eq:newsigma}
\Sigma = e^{i \bm \sigma \cdot \bm \varphi} = \cos \rho+ i \bm \sigma \cdot\hat{\bm \varphi} \sin\rho\,,
\ee
where  $\bm \varphi = \rho \hvf$, thus  $\rho$ describes the  { radial} field, and $\hvf$ is a unit vector field,  such that  $\hvf \cdot \hvf =1$. 
This field decomposition can be thought as  obtained from Eq.~\eqref{eq:barsigma} by promoting $\alpha$ and $\bm n$ to dynamical fields.  In the following we will neglect the $\hf_3$ field, corresponding to the $\pi_0$ direction. We will comment on  the $\pi_0$ field in the Appendix B.   Since we are restricting our analysis to the charged fields,  $\bm n$ is a unit vector in a $2$ dimensional space and  the field $\hvf$ corresponds to only one independent degree of freedom describing the fluctuations of this unit vector. For later convenience we define
\be\label{eq:sqrt_phi_2}
\hat\varphi_1=\sqrt{1-\hat\varphi_2^2}\,,
\ee
meaning that we will treat $\hat\varphi_2$ as the independent degree of freedom. As we will see,  this fluctuation can be identified with the NGB of the broken phase.   We notice that the present procedure resembles closely the one exploited to derive low-energy effective models of $(1+1)d$ quantum magnets, such as the Heisenberg-type ones~\cite{haldane1983a,haldane1983b,affleck1984,haldane1985,affleckLH}. The reason behind this analogy is that although pions are pseudoscalar particles, they have isospin $I=1$. In particular, the charged pions have $I_3=\pm 1$ and  are therefore analogues to magnets in isospin space. Thus, the system can be thought as an isospin quantum magnet and the broken phase corresponds to an ordered magnetic phase, with quantum isospins aligned along one particular direction in isospin space.

Upon substituting Eq.~\eqref{eq:newsigma} in Eq.~\eqref{eq:Lagrangian1}, we obtain the ${\cal O}(p^2)$ Lagrangian 
\begin{align}
\label{eq:lag-rho-phi}
{\cal L}&= \frac{f_\pi^2}{2} \(\partial^\mu \rho\partial_\mu \rho + \sin^2\rho \; \partial^\mu \hf_i\partial_\mu\hf_i \right.\nonumber\\ &\left. -2 m_\pi \gamma \sin^2\rho  \; \epsilon_{3i k} \hf_i \partial_0 \hf_k \) -V(\rho)\,,
\end{align}
where
\be\label{eq:V}
V(\rho) = -f_\pi^2 m_\pi^2\(\cos\rho + \frac{\gamma^2}{2}\sin^2\rho\)\,,
\ee
is the potential. Remarkably, this expression of the ${\cal O}(p^2)$ Lagrangian is not obtained by making any expansion in the fields. The potential does not depend on the field $\hvf$, showing that, at least at this order, the field  $\hvf$ does not have non-derivative couplings, as appropriate for NGBs. Clearly, this result is due to the residual $O(2)$ symmetry  corresponding to the flat direction of the potential.  

The stationary point of the potential can be obtained by solving
\be
\label{eq:rhobar}
\left.\frac{\partial V}{\partial \rho}\right\vert_{\bar\rho} = 0 \qquad \rightarrow  \qquad \bar\rho = \arccos\frac{1}{\gamma^2}\,,
\ee
with $\bar\rho$ corresponding to a minimum of $V$ for $\gamma >1$. Upon substituting this expression in Eq.~\eqref{eq:V} one obtains the same expression of the  pressure reported in Eq.~\eqref{eq:pressure-density}.  On the other hand, for $\gamma <1$ the minimum of the potential is in $\rho =0$. Therefore, the  $\rho$ field has the typical behavior of   the radial mode, which acquires a nontrivial vev in the broken phase. The field $\hf_2$ corresponds to a massless fluctuation, {that is to} the NGB, only in the broken phase when $\sin\bar\rho \neq 0$; see the Appendix B. To further clarify how the massive  and massless modes appear we expand  the  Lagrangian in Eq.~\eqref{eq:lag-rho-phi} close to the stationary point.

To tackle the properties of the NGB boson, we first neglect the radial fluctuations. In the broken phase the Lagrangian for the $\hvf$ field turns out to be
\be
{\cal L}_{\hvf}=\frac{f_\pi^2 \sin^2\bar\rho}2 (\partial^\mu \hf_i\partial_\mu\hf_i  -2 m_\pi \gamma  \epsilon_{3i k} \hf_i \partial_0 \hf_k)\,,
\ee
and to make contact with the standard expression of the NGB Lagrangian we can make a further variable change, 
\be\label{eq:theta-expansion}
\theta= \arctan\(\frac{\hf_2}{\hf_1}\) =   \hf_2 +\frac{2 \hf_2^3}{3}+ {\cal O}(\hf_2^5)\,,
\ee
where in the last equality we have used Eq.~\eqref{eq:sqrt_phi_2} and conveniently assumed that the condensate is oriented along the $1$-direction in isospin space, thus  $\hf_2$ is a small fluctuation. The soft Lagrangian now reads
\be
{\cal L} = \frac{f_\pi^2 \sin^2\bar\rho}2 \partial^\mu \theta \partial_\mu\theta \,,
\ee
where we have neglected  a total derivative. The field $\theta$ is the genuine NGB, or phonon, or Anderson-Bogoliubov mode, because it is  the phase associated with the rotation of the condensate. 
Note that the propagation velocity  of this mode is not equal to the sound speed $c_s$ reported in Eq.~\eqref{eq:cs}, but it is equal to $1$. This result depends on the fact that we have completely neglected the interaction of the phonon with the background,  or more precisely,  with  the radial fluctuations. 

For considering the radial fluctuation in the broken phase we define $\rho = \bar\rho +\chi$, finding that, for $ \gamma \ge1$,
\begin{align}\label{eq:vchi}
V(\chi) =  \frac{f_\pi^2 m_\pi^2}{\gamma^2}\(-\frac{1+\gamma^4}{2} +\frac{\gamma^4-1}{2} \chi^2    \)+{\cal O}(\chi^3)\,,
\end{align}
which shows that  the Higgs mode has a nonnegative mass that vanishes at the phase transition point, as appropriate for second order phase transitions.  

The Lagrangian that includes both the quadratic fluctuations can be obtained from Eqs.~\eqref{eq:lag-rho-phi} and~
\eqref{eq:vchi}, leading to
\begin{align}\label{eq:Lchiphi}
{\cal L} &= -\frac{1}{2} \chi D^{-1} \chi - J \chi + \frac{f_\pi^2 \sin^2 \bar\rho}{2} \partial_\mu \hvf \partial^\mu \hvf +  {\cal L}_\text{s}\,, 
\end{align}
where 
\begin{align}
D^{-1} &= f_\pi^2\(\Box + m_\pi^2\frac{\gamma^4-1}{\gamma^2}\)\,,\\
J & = f_\pi^2\mu_I \sin(2 \bar \rho) \epsilon_{3ij} \hf_i \partial_0\hf_j \label{eq:J}\,,
\end{align}
and 
\be
{\cal L}_\text{s} = -f_\pi^2 \mu_I \sin^2(\bar \rho) (\hf_1 \partial_0\hf_2-\hf_2 \partial_0\hf_1)\,,
\ee
corresponds to a surface term. Indeed, using Eq.~\eqref{eq:sqrt_phi_2}
 this term can be written as a total derivative 
\be
{\cal L}_\text{s} =  -f_\pi^2 \mu_I \sin^2\bar \rho\, \partial_0\arcsin(\hf_2)\,.
\ee
Integrating out the radial fluctuations we obtain the  ${\cal O} (p^2)$ effective Lagrangian describing the propagation and the interaction terms
\begin{align}
{\cal L}_\text{eff}& = {f_\pi^2 \sin^2 \bar\rho}\( \frac{1}{2}\partial_\mu \hf_i \partial^\mu \hf_i \right. \nonumber \\ &+\left.\frac{2}{\gamma^4-1} ((\hf_1 \partial_0\hf_2-\hf_2 \partial_0\hf_1)^2) \)\,,
\end{align}
which we can expand using Eq.~\eqref{eq:sqrt_phi_2} to obtain 
\be
{\cal L}_\text{eff} = \frac{f_\pi^2}{2}\frac{ \gamma^4+3}{\gamma^4} \sum_{n=0}^\infty \hf_2^{2 n}\((\partial_0\hf_2)^2-c_s^2 ({\bm \nabla}\hf_2)^2\)\,,
\ee
with the speed of sound  given by the same expression reported in Eq~\eqref{eq:cs}. It is possible to write the effective Lagrangian in the slightly more compact and suggestive way 
\be\label{eq:Leff}
{\cal L}_\text{eff} = \frac{f_\pi^2}{2}\frac{ \gamma^4+3}{\gamma^4} \sum_{n=0}^\infty \hf_2^{2 n} g^{\mu\nu}\partial_\mu\hf_2\partial_\nu\hf_2\,,
\ee
where $g^{\mu\nu} = \text{diag}(1,-c_s^2,-c_s^2,-c_s^2)$ is the so-called acoustic metric, see for example the discussion in~\cite{Mannarelli:2008jq} and the review \cite{Barcelo:2005fc}.
A remarkable aspect is that the above Lagrangian not only includes the kinetic term but also  all the  ${\cal O} (p^2)$  interaction terms. It is strictly valid for momenta much below $\mu_I$, see  the Appendix B,  and the appearance of the  acoustic metric in Eq.~\eqref{eq:Leff} is due to the fact that the $\hf_2$ field is the analogous of a sound mode.

Using Eq.~\eqref{eq:theta-expansion}, we obtain the Lagrangian for the $\theta$ field
\be
{\cal L} = \frac{f_\pi^2}{2}\frac{\gamma^4 +3}{\gamma^4 } g^{\mu\nu} \partial_\mu \theta\partial_\nu \theta \(1 -3 \theta^2 + {\cal O}(\theta^4)\) \,,
\ee
showing that now this field has the correct propagation velocity. Higher order terms can be obtained by considering higher order terms in Eq.~\eqref{eq:theta-expansion}.

The proposed procedure, relying on the integration of the radial fluctuations, allows us to easily evaluate the kinetic term and the interaction terms of the $\hf_2$ mode, with the additional benefit of having a Lorentz covariant Lagrangian with the effective metric $g^{\mu \nu}$. In principle, in the broken phase it is  possible to diagonalize the Lagrangian, as done for example in \cite{Kogut:2001id,Mammarella:2015pxa}, but this procedure is unnecessarily complicated if one is only interested in momenta below $\mu_I$.

\section{NLO corrections}\label{sec:NLO}
We now explore a different issue. An important result obtained using  the LO chiral Lagrangian is that there is a second order  phase transition between the normal phase and the $\pi$c phase. This means, among other things, that the chiral condensate is continuously rotated into the pion one, see Eqs.~\eqref{eq:chiral_cond} and \eqref{eq:pion_cond}. However, the fact that the rotation is continuous does not seem to  rely on any physical reason, meaning that an  abrupt tilting in some direction in isospin space by an increasing $\mu_I$ would have been a plausible \textit{a priori} possibility. Consider, for example, that the phase transition between the kaon condensed and the pion condensed phases triggered by an increasing $\mu_I$ is of the first order~\cite{Kogut:2001id}.

To study the robustness of the second-order transition we include NLO $\chi$PT corrections. Following \cite{Scherer:2002tk}, for the three-flavor case they are given by
\begin{align}
\label{eq:LOp4}
\CL_4 & = L_1 \left\{\mbox{Tr}[D_{\mu}\Sigma (D^{\mu}\Sigma)^{\dagger}] \right\}^2\nonumber \\ & + L_2 \mbox{Tr} \left [D_{\mu}\Sigma (D_{\nu}\Sigma)^{\dagger}\right] \mbox{Tr} \left [D^{\mu}\Sigma (D^{\nu}\Sigma)^{\dagger} \right] \nonumber\\
&  + L_3 \mbox{Tr}\left[ D_{\mu}\Sigma (D^{\mu}\Sigma)^{\dagger}D_{\nu}\Sigma (D^{\nu}\Sigma)^{\dagger} \right ]\nonumber \\ &+  L_4 \mbox{Tr} \left [ D_{\mu}\Sigma (D^{\mu}\Sigma)^{\dagger} \right ] \mbox{Tr}  \(\chi \Sigma^\dagger+  \chi^\dagger \Sigma\)
\nonumber \\
&  +L_5  \mbox{Tr} \left[ D_{\mu}\Sigma (D^{\mu}\Sigma)^{\dagger}  \(\chi \Sigma^\dagger+  \chi^\dagger \Sigma\)\right] \nonumber \\ &+ L_6  \left[ \mbox{Tr} \(\chi \Sigma^\dagger+  \chi^\dagger \Sigma\) \right]^2   + L_7  \left[ \mbox{Tr}  \(\chi \Sigma^\dagger- \chi^\dagger \Sigma\)\right]^2 \nonumber \\ & + L_8 \mbox{Tr} \left (   \Sigma \chi^\dagger \Sigma \chi^\dagger  +  \chi\Sigma^{\dagger}\chi\Sigma^{\dagger} \right ) + H_2 \mbox{Tr}(\chi\chi^\dagger)\,, 
\end{align}
where $L_i$, with $i=1,\dots,8$, and $H_2$ are the relevant LECs encoding properties of the underlying quark theory.   In principle one can use these NLO corrections for studying the robustness of any phase transition of the $SU(3) $ phase diagram of~\cite{Kogut:2002zg}. Nevertheless, in the present paper we focus on the phase transition between the normal phase and the $\pi_c$ condensed phase. In this case $\chi = m_\pi^2 \mathbf{1}_{2\times2}$ and $\Sigma$ is an $SU(2)$ matrix.  Within this restriction, our $\CL_4$ reduces to the $SU(2)$ expression 
originally derived in~\cite{Gasser:1983yg}, and the $SU(3)$ LECs can be easily mapped to the $SU(2)$ LECs~(see for example \cite{Gasser:1984gg}).

 Upon substituting  Eq.~\eqref{eq:barsigma}  in ${\cal L}_4$ we obtain the following expression for the static NLO Lagrangian
\begin{align}\label{eq:lagrangian_adimensinal}
\bar\CL_\text{stat}^\text{NLO}  &=  f_\pi^2 m_\pi^2 \Big(\cos\alpha+ \frac{\gamma^2}{2}  \sin^2\alpha+  2  a \gamma^4 \sin^4\alpha   \nonumber\\
& + 4  b \gamma^2  \sin^2\alpha \cos\alpha  
+  8  c   \cos^2\alpha +2  d \Big)\,,
\end{align}
where
\begin{align}
a &= \frac{m_\pi^2}{f_\pi^2}(2 L_1 + 2 L_2 + L_3)    \,,  &b =  \frac{m_\pi^2}{f_\pi^2}( 2L_4 + L_5)\nonumber\,, \\ c &= \frac{m_\pi^2}{f_\pi^2} (2 L_6 + L_8)\,,  &d=  \frac{m_\pi^2}{f_\pi^2}(H_2-2 L_8)\label{eq:c_e_d}\,,
 \end{align}
are the relevant combinations of  LECs and  where $\gamma$ is given in Eq.~\eqref{eq:gamma}.
Then, by taking $\alpha=0$ we obtain
\be\label{eq:P0}
P_0 = f_\pi^2 m_\pi^2 \left(1+ 8 c + 2 d\right)\,,
\ee
the NLO expression of  the pressure in the normal phase. At the same order we obtain in the normal phase
\begin{align}\label{eq:mpi4}
m_{\pi,4}^2 &= m_\pi^2 (1+16 c -8 b)\,,  \\
f_{\pi,4}^2 &= f_\pi^2 (1+8 b)\,,
\end{align}
corresponding to the standard ${\cal O}(p^4)$ values of the pion mass and of the pion decay constant, respectively, see for example~\cite{Gasser:1984gg}. 
We remark that $P_0 \neq m_{\pi,4}^2 f_{\pi,4}^2  $, meaning that the pressure renormalization must be carefully taken into account. If one restricts the analysis to the normal phase, this renormalization is  immaterial, because we can subtract an arbitrary constant. However, it is important  to  take into account the nontrivial renormalization of the pressure  $P=P_{\pi c} - P_0$, obtained by subtracting the vacuum pressure to the one of the  $\pi c$ phase. In particular,  the NLO corrections to the pressure of the broken phase depend on $a$, $b$ and $c$ in a nontrivial way and they can   change the value of $\gamma$ for which the transition to the BEC phase takes place. We  refer to $\gamma_c$ as the value of the  $\gamma$ parameter at the phase transition point. 

To make the discussion as simple as possible, it is convenient to subtract the normal phase pressure from Eq.~\eqref{eq:lagrangian_adimensinal}, obtaining the normalized NLO static Lagrangian 
\begin{align}\label{eq:NLOlag}
\CL_\text{stat}^\text{NLO} &=  f_\pi^2 m_\pi^2 (1-z)\Bigg( \frac{\gamma^2}{2}  (1+z)    \nonumber 
  +  2  a \gamma^4 (1-z^2)(1+z)  \\ 
  & - 1 + 4  b\, \gamma^2  z(1+z)-  
 8  c   (1+z)  \Bigg)\,, 
\end{align}
where $ z=\cos\alpha$. The  stationary condition for the Lagrangian is obtained by solving  the cubic equation 
\be
\gamma^2 z -1 + 8  a \gamma^4 (1-z^2) z + 4  b \gamma^2 (3 z^2-1)-16 c z = 0\,,
\ee
considering the appropriate value of $\gamma$. The first order phase transition is obtained when two roots are equal and satisfy $0 \le z \le1$, corresponding to 
\be\label{eq:gammac}
8   a \gamma_c^4 -12  b\,\gamma_c^2 + 8  c > \frac{\gamma_c^2}{2}\,.
\ee
Before discussing the phase transition in  detail, let us notice that for a proper description of the critical value of the isospin chemical potential, one should consider the NLO corrections to the pion mass as well, meaning that the order parameter should be rescaled as follows
\be\label{eq:gamma_rescaled}
\gamma \to \gamma^R \equiv \gamma\frac{m_\pi}{m_{\pi,4}}\,,
\ee 
with the ${\cal O}(p^4)$ rescaled pion mass  defined in  Eq.~\eqref{eq:mpi4}.

\begin{figure}[th!]
\centering
\resizebox{.45\textwidth}{!}{%
\includegraphics{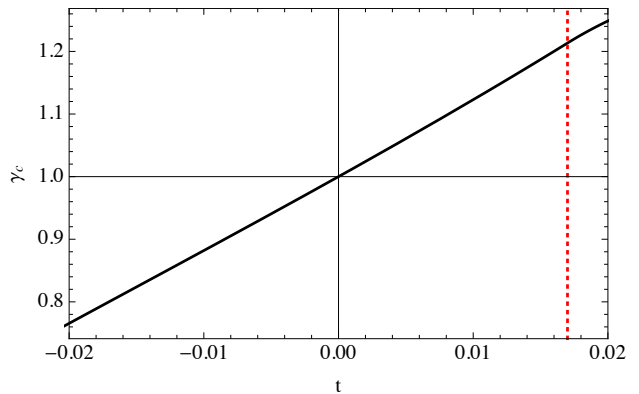}}
\caption{Effect of the NLO $\chi$PT terms on the critical value  for the phase transition between the normal phase and the $\pi$c phase. The NLO effects are parametrized by $t=a=-b=c$, where $a, b$ and $c$ are the combinations of LECs  given in Eq.~\eqref{eq:c_e_d}. The solid black line corresponds to the critical value of $\gamma=\mu_I/m_\pi$.  The dotted vertical red line indicates the onset of the first order phase transition.}
\label{fig:gammac}
\end{figure}

We note at this point that in spite of recent progress with the help of lattice QCD (see \cite{Bijnens:2014lea,Aoki:2013ldr} for recent reviews), some of the LECs are still poorly known. Furthermore, the effects of chiral logarithms, which carry a scale dependence, are often non-negligible \cite{Gasser:1984gg,Bijnens:2014lea}.
Using the values from one of the fits reported in \cite{Bijnens:2014lea} for the LECs evaluated at the scale of the $\rho$ mass, we obtain the following values for our three relevant combinations:
\begin{align}
\label{eq:abcnum}
a&\simeq -0.9 \times 10^{-3}\,, \nonumber \\ b&\simeq -0.9  \times 10^{-3} \,, \nonumber \\ c &\simeq +1.7  \times 10^{-3}\,, 
\end{align}
where we have used $m_\pi=140 $ MeV and $f_\pi=92$ MeV.

Using these values, we find that the NLO corrections shift the second-order transition to $\gamma_c \simeq 0.99$. However, when rescaling the control parameter we obtain  $\gamma_c^R  \simeq 1$.
\begin{figure*}[th!]
\centering
\resizebox{1.\textwidth}{!}{%
\includegraphics{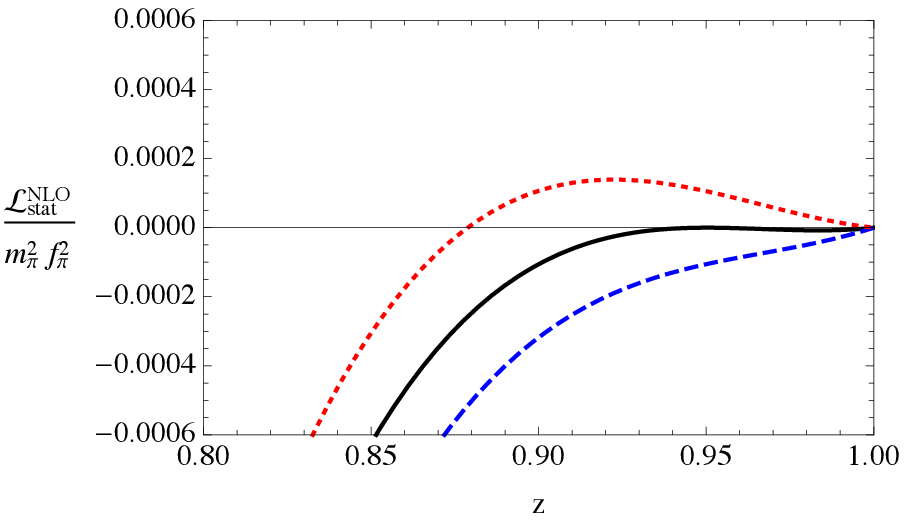}
\includegraphics{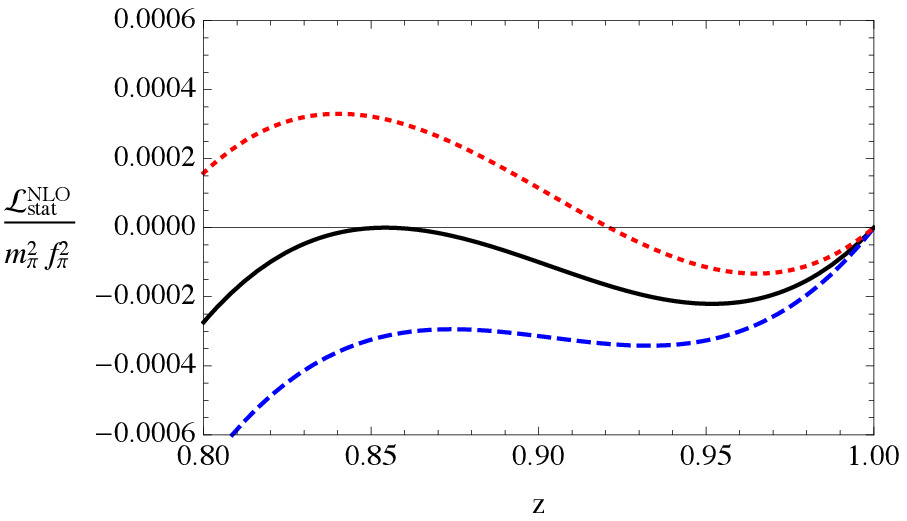}
}
\caption{Lagrangian density, {  given in Eq.~\eqref{eq:NLOlag}}, as a function of the variational parameter $z=\cos\alpha$ considering two different values of $t$, see Eq.~\eqref{eq:t}.  The different lines  correspond to  values of $\gamma$ close to the phase transition between the normal phase and the pion condensed  phase. { To be more specific,} the dotted (red) lines correspond to $\gamma=\gamma_c+0.001$, the solid (black) lines correspond to $\gamma=\gamma_c$ and the dashed (blue) lines have been obtained with $\gamma=\gamma_c-0.001$. The maximum of ${\cal L}_\text{stat}^\text{NLO}$ corresponds to the NLO pressure of the system. Left panel: results obtained taking $t=0.018$,   the first order phase transition leads  to a change $\Delta z\simeq 0.05$ of the tilting angle of the pion condensate. Right panel: results obtained taking $t=0.02$,  the first order phase transition corresponds to a change  $\Delta z\simeq 0.15$ of the pion condensate.}
\label{fig:pressure}
\end{figure*}

Let us now explore what happens for different values of the NLO parameters. Since $b$ enters with a minus sign in Eq.~\eqref{eq:gammac} and to simplify the analysis, we limit ourselves to a particular parameter subspace characterized by 
\be\label{eq:t} t=a=-b=c\,. \ee
Clearly, a more refined analysis can be done, however, especially considering the uncertainty on the values of the LECs,  it seems to us more appropriate to conduct a qualitative study in terms of one single variable.

In Fig.~\ref{fig:gammac} we report the value of $\gamma_c$ (solid black line) as a function of $t$. The vertical dotted line corresponds to the onset of the first order phase transition: the second order phase transition turns into a first order phase transition for $t\gtrsim 0.017$. We note that depending on the sign of $t$, $\gamma_c$ can be larger or smaller than $1$, however $\gamma_c^R \simeq 1$ for any considered value of $t$. This suggests that the  phase transition happens at $\mu_I = m_{\pi,4}$.

In order to better understand the effect of the  NLO corrections on the  order of the phase transition,   we report in Fig.~\ref{fig:pressure} the plot of the static Lagrangian for $t=0.018$, left panel, and for $t=0.02$, right panel. We consider three values of $\gamma$: below the phase transition point (dashed line), at the phase transition point (solid line) and in the pion condensed phase (dotted line).   { Close to  the phase transition the   vacuum  and the LO pressures balance. Therefore the quantity reported in Fig.~\ref{fig:pressure} basically corresponds to the NLO contribution,  generating a  small additional structure  inducing a weak first-order transition.}  The pressure of the system corresponds to the maximum of the static Lagrangian and  is obviously a continuous function of $\gamma$. However, the tilting angle $\alpha$ of the ground state, see Eq.~\eqref{eq:barsigma},  is discontinuous. This is accompanied by a discontinuity in
  the number density  with a jump $\Delta n_I \propto \sqrt{\Delta z}$. Present lattice QCD simulations, see~\cite{Detmold:2012wc}, use $m_\pi=390$ MeV, but do not seem to show any jump in $n_I$ at the phase transition point.  It might be of interest to implement  lattice QCD simulations with a larger value of the pion mass. In this case the physical value of the phase transition point does not change much because $\gamma_c^R\simeq 1$, however the phase transition might turn to be of the first order type, because a larger pion mass implies a larger $t$.

\section{Conclusions}\label{sec:conclusion}
The properties of matter at nonvanishing isospin chemical potential are quite interesting and in some aspects highly non-trivial. We have shown that the formation of the pion condensate can be described by a standard GP Lagrangian, in which all the coefficients depend on $\mu_I$. This is one of the reasons why the description of the condensation mechanism looks complicated. The GP approximation breaks down deep in the pion condensed phase, because it corresponds to a low-density approximation and the number density of pions in the ground state grows with $\mu_I$. For this reason we have obtained a different expansion of the $\chi$PT Lagrangian, in which the identification of the massless NGB is more direct than in the standard approach. This new approach leads to a Lagrangian similar to the one used for describing quantum magnets, with the isospin playing the role of the spin in condensed matter systems. The broken phase can be identified with an ordered  magnetic phase in which the isospins are aligned along the  direction of the isospin chemical potential. Indeed, the isospin chemical potential enters in the chiral Lagrangian as an external source, pretty much as the magnetic field enters in the Lagrangian of  a quantum magnet.  A positive chemical potential leads to the alignment of $\pi_+$ mesons, that corresponds to a $\langle \pi_+ \rangle$ condensate. On the other hand, a negative chemical potential aligns the $\pi_-$ mesons, corresponding to a $\langle \pi_-\rangle$ condensate. We have obtained all the interaction  and surface terms of the soft Lagrangian by integrating out the radial fluctuations around the vev. The  kinetic and interaction terms can be written in the compact form reported in Eq.~\eqref{eq:Leff}, in which we made use of the acoustic metric emerging from the interaction of the NGB with the vacuum fluctuations.

Finally, we have tested the order of the phase transition between the normal phase and the pion condensed phase including NLO chiral corrections. For standard values of the LECs   the transition remains of the second order, but an intriguing possibility is that lattice QCD simulations with a very large pion mass might observe a first order phase transition. Moreover, more refined lattice QCD simulations could be used to determine the combinations of the LECs that appear in the NLO chiral Lagrangian.

\begin{appendix}
\section*{Appendix A: Self-bound pion stars}\label{sec:Pion stars}
Since in the $\pi c$ phase one of the charged pions is stable~\cite{Mammarella:2015pxa},   we speculate on the possible existence of stars consisting of condensed pions. We will assume  a weak first order phase transition between the normal phase and the $\pi c$ phase, meaning that we focus on the region $t \gtrsim 0.017$, see Sec.~\ref{sec:NLO},  probably corresponding to  unphysical values for the LECs. 

This  system, if sufficiently cold, will be   self-bound and it would not spread as a gas even if it is so small to be 
gravitationally unbound. The resulting stellar object would be a particular type of Bose star. Bose stars are stellar objects consisting of a large number of bosons in which the boson wave-function varies inside the star, see for example~\cite{Lee:1991ax}.  For noninteracting bosons with mass $m$, the maximum mass of a Bose star has been determined in~\cite{Breit:1983nr} and turns out to be $M_\text{max} = 0.633/(Gm)$, where $G$ is the gravitational constant. For self-interacting bosons, the maximum mass can be larger, as  shown in~\cite{Braaten:2015eeu} for axion stars. Certainly, Bose stars can only exist if a stable boson  exists.

Let us  assume that  matter in the $\pi$c phase is produced by some astrophysical event. Since it is self-bound it will not evaporate and can possibly accrete matter capturing positrons, converting them in $\pi_+$ and ejecting neutrinos.  In other words,  a small number of pions  produced with a certain asymmetric mechanism in such a way that there are, say, much more $\pi_+$ than $\pi_-$ (meaning that  $\mu_I>0$) may become larger and larger if the $\pi_+$ condense. Indeed, if  this chunk of matter is sufficiently cold,  the  $\pi_+$ will form a condensate and will not decay in leptons~\cite{Mammarella:2015pxa}. On the other hand the $\pi_-$ do not condense and 
will quickly decay, mainly in muons and corresponding neutrinos~\cite{Mammarella:2015pxa}.  However, we expect that the system does not eject electrically charged particles because the electrons produced by the muon decay  will be  bound by the electromagnetic force.  
 
The neutrality condition for this peculiar system reads
\be n_I = n_e\,,\ee
where $n_e$ is the electron number density. { Although the isospin chemical potential and the electron chemical potential are not equal,  a relation between them  can be obtained once the number densities are expressed in terms of the corresponding chemical potentials.}   At $T=0$ the process $ e^- + \pi_+ \to \nu_e $ cannot happen because all pions are in the condensate, corresponding to a state of zero momentum. However, at $T>0$, part of the $\pi_+$ are in excited states and can be annihilated by electrons in neutrinos. These neutrinos will certainly escape leading to a cooling of the stellar object and a related reduction of the excited $\pi_+$. For this reason    we can conveniently  consider the $T=0$ case, meaning that no quasiparticle thermal excitation is present.

If the system is sufficiently big it can certainly gravitationally capture neutral particles, like neutrons or atoms, but in the following we will assume the simplified scenario that it only consists of  $e^-$ and condensed $\pi_+$.

Solving the Tolman-Oppenheimer-Volkoff equation, see for example~\cite{Shapiro-Teukolsky}, one can determine the mass-radius sequence for a system of pions in the $\pi$c phase neutralized by a gas of electrons.  Since matter is self-bound we expect  a mass-radius trajectory similar to that of strange-stars~\cite{Alcock:1986hz}, see for example~\cite{Mannarelli:2014ija}. The maximum mass of these stellar objects depends on the various parameters of the pion pressure. As a preliminary result we have found  that using  $m_\pi=140$ MeV, $f_\pi=92$ MeV and  $t=0.02$, stellar masses up to few times the solar mass with radii of tens of kilometers can be formed. Further analysis is ongoing and we expect to report on this topic in the future.

\section*{Appendix B: More about the alternative description of the $\pi c$ phase}\label{sec:Aspects}

Let us clarify some aspects of the procedures presented in Sec.~\ref{sec:alternative}  that might be of some concern.

First of all, in the expansion of the potential in  Eq.~\eqref{eq:vchi} we have neglected terms of order $\chi^3$ and $\chi^2 \partial_0\hf_2$. To make sure that our approach is correct, we derive from Eq.~\eqref{eq:Lchiphi}  the equation of motion of the $\chi$ field,  
\be
\(\Box + m_\pi^2\frac{\gamma^4-1}{\gamma^2}\) \chi = \mu_I \sin(2 \bar \rho) (\hf_1 \partial_0\hf_2-\hf_2 \partial_0\hf_1)\,,
\ee 
that we can rewrite using Eq.~\eqref{eq:sqrt_phi_2} as
\be
\(\Box + m_\pi^2\frac{\gamma^4-1}{\gamma^2}\) \chi = \mu_I \sin(2 \bar \rho) \partial_0\arcsin(\hf_2)\,,
\ee 
meaning that for any $\gamma>1$ and at the leading derivative order,  $\chi \propto \partial_0\hf_2$.  Therefore both $\chi^3$ and  $\chi^2 \partial_0\hf_2$ terms are ${\cal O}(p^3)$ and should not  be included in the ${\cal O}(p^2)$ Lagrangian.

The second aspect that deserves some comment is related to the fate of the $\pi_0$ field. In Sec.~\ref{sec:alternative} we have neglected this  mode, but it can be  straightforwardly included by considering $\hvf=(\hf_1,\hf_2,\hf_3)$.  This mode does not couple with the radial fluctuations, because of the Levi-Civita symbol in Eq.~\eqref{eq:J}. Therefore, it does not feel the background fluctuations leading to the renormalization of the propagation speed. In other words, the Lagrangian of this field is simply given by a standard Lorentz invariant expression. In writing the Lagrangian including this mode it is important to consider that  the potential term in Eq.~\eqref{eq:V} must be written in the appropriate way
\be
\label{eq:V3pi}
V = -f_\pi^2 m_\pi^2\(\cos\rho + (\hf_1^2+\hf_2^2) \frac{\gamma^2}{2}\sin^2\rho\)\,,
\ee
because now $\hf_1^2+\hf_2^2 \neq 1$.
In both the normal phase and the broken phase we can write the quadratic Lagrangian in the separable form
\be
{\cal L} = {\cal L}_{\varphi_1,\varphi_2} + {\cal L}_{\varphi_3}\,,
\ee
thus we can restrict, at the quadratic order, to considering the ${\cal L}_{\varphi_1,\varphi_2}$ Lagrangian. At higher orders in the fields this separation is no longer possible. However, in the broken phase  $m_{\pi^0} = \mu_I$, therefore, this mode decouples from the soft-Lagrangian when considering momenta and energies below this scale. For this reason  the soft Lagrangian in  Eq.~\eqref{eq:Leff} is  valid for momenta much smaller than $\mu_I$.

Finally, let us sketch how the standard expression of the quadratic Lagrangian, as given for example in~\cite{Mammarella:2015pxa}, can be obtained starting from the expression in terms of radial and angular fields. We focus on  the normal phase (in the broken phase  a similar reasoning can be used).   The leading order Lagrangian  is given by
\be\label{eq:Lnorm}
{\cal L}=\frac{f_\pi^2}{2}(\partial^\mu\chi \partial_\mu\chi + \chi^2 \partial^\mu\hf_i \partial_\mu\hf_i - 2 m_\pi \gamma \chi^2 \epsilon_{3ij}\hf_i\partial_0\hf_j) - V(\chi)\,
\ee
with the potential term in Eq.~\eqref{eq:V3pi}. By definition  $\varphi_i=\chi \hf_i$, then considering that 
\be
\partial^\mu\chi \partial_\mu\chi + \chi^2 \partial^\mu\hf_i \partial_\mu\hf_i = \partial^\mu\varphi_i \partial_\mu\varphi_i\,,
\ee
we obtain the quadratic Lagrangian 
\begin{align}
\label{eq:Lnormphi}
{\cal L}&=\frac{f_\pi^2}{2}\( \partial^\mu\varphi_i \partial_\mu\varphi_i-2 m_\pi\gamma  \epsilon_{3 i j} \varphi_i \partial_0 \varphi_j \) \nonumber \\&+ f_\pi^2m_\pi^2 \(1- \frac{\varphi_3^2}{2} - \frac{1-\gamma^2}{2}(\varphi_1^2 +\varphi_2^2 )\) \,,
\end{align}
that is the standard expression for the $O(p^2)$ quadratic Lagrangian. 
Note that  in the normal phase one can distinguish three modes,  but only one mode appears in the expression of the Lagrangian in Eq.~\eqref{eq:Leff}. It seems, therefore, that two modes disappear in the transition from the normal phase to the broken phase. As discussed above, one of this mode is related to the $\pi_0$, that decouples. The second mode that disappears is actually integrated out. Indeed, at the phase transition point the radial mode develops a nontrivial vev and, as shown in Sec.~\ref{sec:alternative}, the small fluctuations around it  can be integrated out. Note that the mode that is integrated out is not one of the two charged modes, instead  it is given by the combination of the charged modes that corresponds to the radial fluctuation.

\end{appendix}

\bibliographystyle{epj}

\end{document}